\documentclass[aps,prl,showpacs,twocolumn]{revtex4}

\newif\ifpdf
\ifx\pdfoutput\undefined \else \pdftrue \fi

\ifpdf
\usepackage[pdftex]{graphicx}
\else
\usepackage{graphicx}
\fi
\usepackage{hyperref}
\usepackage{amssymb}
\usepackage{amsfonts}

\begin{document}

\ifpdf \DeclareGraphicsExtensions{.pdf, .jpg} \else
\DeclareGraphicsExtensions{.eps, .jpg} \fi

\def\hslash{\hbar}
\def\imag{i}
\def\grad{\vec{\nabla}}
\def\div{\vec{\nabla}\cdot}
\def\curl{\vec{\nabla}\times}
\def\DDt{\frac{d}{dt}}
\def\ddt{\frac{\partial}{\partial t}}
\def\ddx{\frac{\partial}{\partial x}}
\def\ddy{\frac{\partial}{\partial y}}
\def\lap{\nabla^{2}}
\def\divv{\vec{\nabla}\cdot\vec{v}}
\def\gradS{\vec{\nabla}S}
\def\vvec{\vec{v}}
\def\wc{\omega_{c}}
\def\<{\langle}
\def\>{\rangle}
\def\Tr{{\rm Tr}}
\def\Csch{{\rm csch}}
\def\Coth{{\rm coth}}
\def\Tanh{{\rm tanh}}
\def\g2{g^{(2)}}

\title{Spin-dependent electron-hole capture kinetics in
conjugated polymers}

\author{Stoyan Karabunarliev}
\email[email:]{karabunarliev@uh.edu} \affiliation{Department of
Chemistry, University of Houston, Houston, TX 77204-5003}
\author{Eric R. Bittner}
\email[email:]{bittner@uh.edu} \affiliation{Department of
Chemistry, University of Houston, Houston, TX 77204-5003}

\date{\today}

\begin{abstract}
The recombination of electron-hole pairs injected in extended
conjugated systems is modeled as a multi-pathway vibron-driven
relaxation in monoexcited state-space. The computed
triplet-to-singlet ratio of exciton formation times  $r =
\tau_T/\tau_S$ increases from 0.9 for a model dimer to 2.5 for a
32-unit chain, in excellent agreement with experiments. Therewith
we rationalize recombination efficiency in terms of spin-dependent
interstate vibronic coupling and spin- and
conjugation-length-dependent exciton binding energies.

\end{abstract}
 \maketitle

{\em Introduction} - The efficiency of electroluminescence
(EL)from conjugated polymers in light emitting diodes (LEDs) is
largely determined by the fraction of injected electron-hole (e-h)
pairs which combine to form emissive spin-singlet ($S$) as opposed
to non-emissive spin-triplet ($T$) bound states.\cite{ref1} If one
assumes that the cross sections for $S$ and $T$ e-h captures are
equivalent $\sigma_S = \sigma_T$, then the singlet generation
fraction $\chi_S = \sigma_S/(\sigma_S+3\sigma_T)$ will be limited
to 25\% according to spin multiplicity.  Nonetheless, efficiencies
corresponding to $\chi_{S}$ over 50\% have been achieved in
poly($p$-phenylenevinylene) (PPV) based LEDs.\cite{ref2,ref3} From
this it has been inferred that singlet e-h capture is
intrinsically more efficient than the respective triplet process,
$\sigma_S > \sigma_T$. Recent PA/PADMR experiments on a wide
variety of conjugated systems indicate that the cross-section
ratio $r = \sigma_S/\sigma_T$ typically lies between 2 and 5,
\cite{ref4} and that the ratio increases with conjugation
length.\cite{ref5}

The observed variation of $r$ was originally related\cite{ref5} to
the optical gap according to an interchain recombination
model\cite{ref6}; however, more recent experiments indicate that
large values of $r$ are characteristic of extended intrachain
conjugation.\cite{ref5,ref7} Wilson {\em et al.} have measured
$\chi_S$ and $\chi_T$ consistent with $r\approx 4$ for a polymer,
but only $r\approx 0.9$ for the monomer.\cite{ref7} While
time-dependent scattering calculations by Kobrak and
Bittner\cite{ref8} indicate that the weakly bound $S_1$ exciton is
a more efficient trap of free e-h pairs than the tightly bound
$T_1$ triplet exciton, the systematic variation of $r$ with
(effective) conjugation length $n$ has remained unclear.

This Letter addresses the kinetics of e-h capture in extended
conjugated systems. To our best knowledge, ours is the first
molecular-based approach which consistently reproduces the spin-
and conjugation-length dependence of the process, and rationalizes
it in terms of exciton binding energies. This has important
ramifications in the design and synthesis of polymer materials for
device applications.

{\em Method} - We simulate the recombination process in
nondegenerate $\pi$-chains by the dissipative dynamics of the
multi-level electronic system coupled to the phonon bath. The
detailed structure of the Hamiltionian with a general form
\begin{eqnarray}
{\hat H} &=& {\hat H}_{el} + {\hat H}_{ph} + {\hat H}_{el-ph}.
\end{eqnarray}
is described in a separate publication,\cite{ref9} and we mention
only those salient features that are relevant herein. The
electronic part ${\hat H}_{el}$ reflects the configuration
interaction (CI) of single excitations for a generic two-band
polymer in a localized basis.\cite{ref10,ref11} The configurations
$|{\bf m}\> = |\overline{m}m\>$ taken over valence and
conduction-band Wannier functions $|\overline{m}\>$ and $|m\>$
represent geminate $(\overline{m} = m)$ and charge-transfer
$(\overline{m}\ne m)$ e-h pairs.  $hat H_{el}$ is parameterized
from the $\pi$-band structure of extended PPV.  Since the
conjugated backbone is alternant, one-body CI integrals $F_{\bf
mn} = \delta_{\overline{m}\overline{n}}f_{m-n} -
\delta_{{m}{n}}\overline{f}_{\overline{m}-\overline{n}}$ of the
band structure operator $f$ obey electron-hole symmetry with
$\overline{f}_r = -f_r$. Spin dependency originates from two-body
interactions with $V_{\bf mn} = -\< m \overline{n}||
n\overline{m}\> + 2\<m\overline{n}||\overline{m}n\>$ for singlets
and $V_{\bf mn} = -\< m\overline{n}|| n\overline{m}\>$ for
triplets.  Assuming zero-differential overlap except for geminate
orbitals, we take into account only true Coulomb and exchange
terms $\<m\overline{m}||m\overline{m}\>$ and
$\<m\overline{m}||\overline{m}m\>$, and (transition) dipole-dipole
interactions $\<m\overline{n}||\overline{m}n\>$ between geminate
singlet pairs.  The electron-phonon coupling term $\hat H_{el-ph}$
assumes that localized conduction/valence levels $\pm f_o$ and
nearest-neighbor transfer integrals $\pm f_1$ are modulated by
lattice oscillations. Correspondingly, the phonon term consists of
two sets of interacting local oscillators, giving rise to two
dispersed optical phonon branches centered at 1600 and
100cm$^{-1}$, respectively.  These frequencies roughly correspond
to the C=C bond stretches and ring torsions in PPV, which largely
dominate the Franck-Condon activity of the lowest optical
transitions.\cite{ref12,ref13} Thus, by empirical adjustment of
el-ph coupling strength in $\hat H_{el-ph}$, we compute in Condon
approximation\cite{ref13} fairly accurate absorption and emission
band-shapes for PPV chains, reflecting conjugation-length
dependence, vibronic structure and Stokes shifts.

Diagonalizing $\hat H_{el}$ and $\hat H_{ph}$ yields vertical
excited states with energies $\varepsilon_a^o$\cite{ref14} and
normal modes with frequencies $\omega_\xi$, which allow us to
transform the el-ph coupling term in its diabatic representation.
\begin{eqnarray}
{\hat H} &=& \sum_a \varepsilon_a^o |{\bf a}\>\<{\bf a}|
   + \frac{1}{2}\sum_{\xi}\left( \omega_\xi^2 Q_{\xi}^2 +
P_\xi^2\right)\nonumber \\
& +& \sum_{a b \xi} g_{ab\xi}^o Q_\xi |{\bf a}\>\<{\bf b}|
.\label{Hdiabatic}
\end{eqnarray}
FIG.1 shows the $S$ and $T$ densities of states (DOS) for a
conjugated chain with $n$ = 32 repeat units (PPV$_{32}$), obtained
as a convolution of Lorentzian line-shapes with $\Gamma$ = 0.01eV.
We note that in singlet state-space, optical coupling with the
ground state is concentrated mainly around the DOS bottom since
only $S_{1}$ and other low-lying excitonic states contain geminate
configurations with transition dipoles in additive
combinations.\cite{ref9}
\begin{figure}[t]
\includegraphics[width=3.4in]{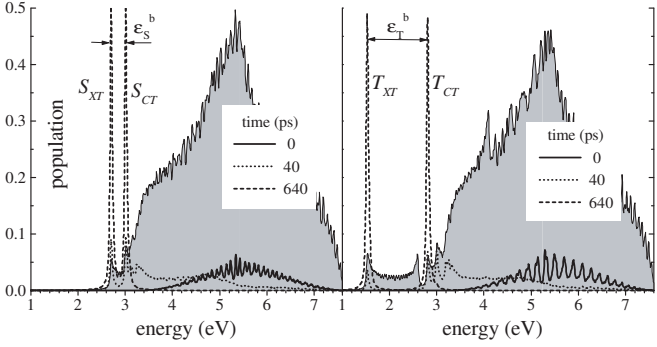}
\caption{Density of spin-singlet (left) and spin-triplet (right)
states for PPV$_{32}$ (shaded area)  and their populations at
different recombination stages.  The $t=0$ population corresponds
to the initially injected e-h pair.}\label{DOS-PPV}
\end{figure}
Nonequilibrium dynamics in excited state-space is derived further
from the diabatic interstate vibronic couplings $g_{ab\xi}^\circ$.
Assuming that phonons thermalize rapidly on the time scale of
electronic relaxation, internal conversion (IC) rates $k_{ab}$ for
one-phonon processes are obtained in Markov
approximation.\cite{ref15}
\begin{eqnarray}
k_{ab} = \pi\sum_{\xi} \frac{(g_{ab\xi}^o)^2}{\hbar\omega_\xi}
(n_\xi + 1) ( \Gamma(\omega_\xi-\omega_{ab})-
\Gamma(\omega_\xi+\omega_{ab}))
\end{eqnarray}
Here $n_\xi$ is the Bose-Einstein distribution of phonons at T =
300K and $\Gamma$ is the empirical broadening.  Note that for an
elementary $a\rightarrow b$ conversion to occur there must a phonon
mode $\omega_\xi$ close to the transition frequency
$\omega_{ab}=(\varepsilon_a^o-\varepsilon_b^o)/\hbar$.  Singlet
and triplet superposition states are separately evolved in time
according to the quantum master equation for the density matrix
$\rho$ in energy representation,
\begin{eqnarray}
\dot \rho_{ab} =-i\omega_{ab}\rho_{ab}-\sum_{cd}
{R}_{ab,cd}\rho_{cd}
\end{eqnarray}
where we decouple populations $\rho_{aa}$  from coherences
$\rho_{ab}$ according to the Bloch model.\cite{ref15}
\begin{eqnarray}
{R}_{aa,bb} = -k_{ab} + \delta_{ab}\sum_c k_{ac}\\
{R}_{ab,ab} = \frac{1}{2}\sum_c (k_{ac}+k_{bc})
\end{eqnarray}
{\em Results} - Equations (3-6) allow us to model the relaxation
of an arbirtary monoexcitation, including that with an electron
and hole injected into the far ends of the conjugated chain. The
resulting initial population of the DOS is shown in FIG.1. We show
as well the populations at an early relaxation time (40ps), and at
640ps when a stationary state is reached. In both spin-spaces,
half of the population density propagates down to the lowest
excitons ($S_{XT}$ or $T_{XT}$),\cite{ref16} but the other half
remains locked in metastable charge-transfer (CT) states ($S_{CT}$
or $T_{CT}$). The branching of the relaxation pathways is due to
e-h symmetry, which separates excited states into even and odd
under e-h transposition. The XT states, which are the lowest even
ones, are not vibronically coupled with the CT states, which the
lowest odd ones. XT and CT states are ultimately populated equally
because the initial separated e-h pair is a 1:1 mixture of
eigenstates with even and odd e-h parity. The weak e-h symmetry
breaking of real conjugated systems is further modeled by slight
conduction/valence band nonsymmetry according to the relation
$f_{r}/\overline{f}_{r}= -1.1$. The perturbation barely affects
the DOS, but invokes weak vibronic coupling between the states of
odd and even types, so that they undergo slow mutual conversions.
The time-dependent XT and CT populations are shown in FIG.2 both
for the symmetric (a) and  nonsymmetric (b) cases. We see that
when e-h symmetry is broken, the fast formation of XT and CT
during the the first 100-200ps is followed up by a slow
CT$\rightarrow$XT relaxation. Most notably, in comparison with the
symmetric case, the initial XT-CT branching ratio changes
drastically in favor of $S_{XT}$ for singlets and in disfavor of
$T_{XT}$ for triplets. The subsequent $S_{CT}\rightarrow S_{XT}$
relaxation is also substantially faster than the respective
$T_{CT}\rightarrow T_{XT}$ conversion process, and $S_{XT}$
population remains about twice as high as that of $T_{XT}$
throughout the simulated time-range. We note here that CT states
are barely spin-dependent (see FIG.1) due to negligible
spin-exchange; hence $S\leftrightarrows T$ intersystem crossing
pre-equilibrium of long-lived CT states is also likely to occur
prior to final e-h binding.
\begin{figure}
\includegraphics[width=3.5in]{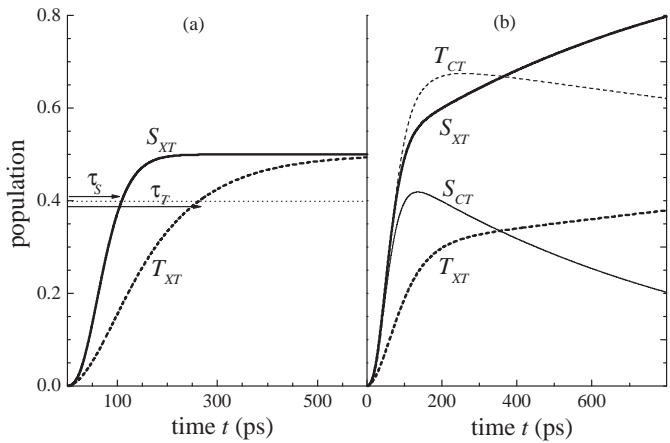}
\caption{Essential-state populations as a function of time for
PPV$_{32}$ with strict (a) and weakly broken (b) e-h symmetry. The
$S_{CT}$ and $T_{CT}$ populations not shown in (a) follow closely
the $S_{XT}$ curve. }\label{rates}
\end{figure}
As shown in FIG.2a, $S_{XT}$ formation proceeds notably faster
than that of $T_{XT}$. Since even under e-h symmetry population
buildup deviates from first-order kinetics in early stages when
intermediate states are formed (see FIG.1), we measure the
relative efficiency of exciton formation by the time $\tau$ at
which XT population reaches 40\%. Thus for PPV$_{32}$ we find
$\tau_{S}$ = 109ps and $\tau_{T}$ = 263ps, which corresponds to $r
= \tau_{T}/\tau_{S}$ = 2.5 or $\chi_{S}$ = 45\% in good agreement
with both EL\cite{ref2,ref3} and PA/PADMR\cite{ref4,ref5} data.
Singlet versus triplet recombination enhancement is related to the
difference in exciton binding energies,\cite{ref17} which we
measure by $\varepsilon^b =
\varepsilon^o_{CT}-\varepsilon^o_{XT}$. Whereas $S_{XT}$ is
slightly lower than $S_{CT}$, spin-exchange contributes
$\approx$1eV more to the triplet binding energy $\varepsilon^b_T$.
Therefore relaxation into $T_{XT}$ takes on the average a longer
sequence of internal conversions, whose energy cutoff at
$\approx$0.2eV is determined by the phonon spectrum. If we assume
that relaxation time is proportional to the dissipated energy and
reciprocal to the mutual coupling of the states along the pathway,
we can write down,
\begin{eqnarray}
r = \frac{\tau_T}{\tau_S}\propto r_{vib}
\frac{\varepsilon^b_T}{\varepsilon^b_S}
\end{eqnarray}
where $r_{vib}$ reflects any difference of the effective vibronic
coupling in $S$ and $T$ state-space.
\begin{figure}
\includegraphics[width=3.4in]{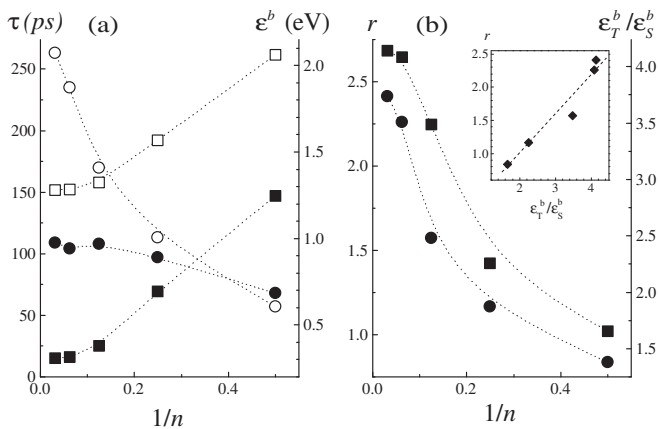}
\caption{ (a) Variation of singlet and triplet binding energies
and formation times with inverse conjugation length $1/n$.
($\tau_S=\bullet$, $\tau_T=\circ$, $\varepsilon^b_S=\blacksquare$,
$\varepsilon^b_S=\square $). (b) Binding energy and formation-time
ratio $\tau_T/\tau_S$ $\bullet$ and binding-energy ratio
$\varepsilon^b_S/\varepsilon^b_T$ $\blacksquare$ vs. $1/n$. Inset:
$r$ vs. $\varepsilon^b_S/\varepsilon^b_T$. Dashed curves are
guides for the eye.}\label{bind}
\end{figure}
We further probe relation (7) by varying the conjugation length
$n$. For the sake of brevity, results for the series of PPV-model
chains with $n$ = 2, 4, 8, 16, and 32 are summarized in FIG.3a,
where exciton binding energies $\varepsilon^b$ and formation times
$\tau$ are plotted vs. inverse chain length $1/n$ as commonly
adopted. Note that $\varepsilon^b_S$ and $\varepsilon^b_T$
decrease in parallel with $1/n$ and converge yet for the longest
chains, suggesting that effective conjugation length is over 10
repeat units. Exciton formation times show different behavior.
Whereas $\tau_{T}$ increases about four times on going from the
dimer to PPV$_{32}$,  $\tau_{S}$ increases only slightly. In FIG.
3b we plot $r$ vs. $1/n$ and establish an excellent agreement with
experiments\cite{ref5,ref7} for both oligomers and polymers,
including a notable triplet enhancement with $r = 0.9$ for $n= 2$.
The inset is a plot of $r$ vs. $\varepsilon^b_T/\varepsilon^b_S$,
which illustrates the existence of a rough linear relationship (7)
with $r_{vib}$ similar to 0.6.

{\em Conclusion} - In summary, we have proposed and modeled an
intrachain e-h recombination mechanism for conjugated systems,
with computational results consistent with the spin-resolved
experiments in all essential points. The model reflects the notion
that exciton formation occurs via vibron-mediated internal
conversions across excited states, and hence the overall rate of
electronic relaxation is inversely proportional to the energy
dissipated into vibrons. Hence while strong vibronic coupling of
the lowest triplet is responsible for $r < 1$ in short chains and
molecular species, the disproportion of exciton binding energies
accounts for the large $r$ values in extended polymers. The
approximate e-h symmetry of conjugated systems implies the
parallel formation of long-lived charge-transfer states, which are
nearly spin-independent, and thus susceptible to intersystem
crossing prior to final spin-singlet enhanced e-h binding.

This work was funded by the National Science Foundation (CAREER
Award) and the Robert A. Welch Foundation.


\begin{thebibliography}{10}

\bibitem{ref1} R. H. Friend et al., Nature {\bf 397}, 121 (1999).

\bibitem{ref2} Y. Cao et al., Nature {\bf 397}, 414 (1999).

\bibitem{ref3} J. S. Kim et al., J. Appl. Phys. {\bf 88}, 1073 (2000).

\bibitem{ref4} M. Wohlgenannt et al., Nature {\bf 409}, 494 (2001).

\bibitem{ref5} M. Wohlgenannt et al., Phys. Rev. Lett. {\bf 88}, 197401
(2002)

\bibitem{ref6} Z. Shuai et al.,  Phys. Rev. Lett. {\bf 84}, 131 (2000);
A. Ye et al., Phys. Rev. B {\bf 65}, 5208 (2002).

\bibitem{ref7} J. S. Wilson et al., Nature {\bf 413},  828 (2001).

\bibitem{ref8} M. Kobrak and E. R. Bittner, Phys. Rev. B {\bf 62}, 11473
(2000).

\bibitem{ref9} S. Karabunarliev and E. R. Bittner, cond-mat/0206015.

\bibitem{ref10} D. Mukhopadhyay et al., Phys. Rev. B {\bf 51}, 9476
(1995); M. Chandross et al., Phys. Rev. B {\bf59}, 4822 (1999).

\bibitem{ref11} P. Karadakov et al., J. Chem. Phys. {\bf94}, 8520
(1991).

\bibitem{ref12} T. W. Hagler et al., Phys. Rev. B {\bf49}, 10968 (1994);
K. Pichler et al., J. Phys. Cond. Mat. {\bf5}, 7155 (1993); J.
Cornil et al., Chem. Phys. Lett. {\bf278}, 139 (1997).

\bibitem{ref13} S. Karabunarliev et al., J. Chem. Phys. {\bf113}, 11372
(2000); {\bf114}, 5863 (2001).

\bibitem{ref14} Superscript $^o$ applies to quantities, taken at
ground-state equilibrium.

\bibitem{ref15} For a recent review see: V. May and O. Kuhn, {\em Charge
and Energy Transfer Dynamics in Molecular Systems} (Wiley-VCH,
Berlin, 2000).

\bibitem{ref16} By $S_{XT}$ we mean $S_1$ and the several other
thermally populated singlet states above it. Similarly for
$S_{CT}$, $T_{XT}$, and $T_{CT}$.

\bibitem{ref17} J. L. Bredas, J. Cornill, and A. J. Heeger, Adv. Mater.
{\bf8}, 447 (1996); J. L. Bredas et al., Acc. Chem. Res. {\bf32},
267 (1999).

\end{thebibliography}
\end{document}